\documentclass[10pt]{IEEEtran}

\usepackage{amsmath,amsfonts,graphicx,float}
\usepackage{xcolor}
\usepackage{multicol}
\usepackage{multirow}
\usepackage[justification=centering]{caption}
\usepackage{subcaption}
\begin{document}
\title{Subjective Assessment of H.264 Compressed Stereoscopic Video}

\author{Manasa K,~Balasubramanyam~Appina,~and~Sumohana~S.~Channappayya,~\IEEEmembership{Member,~IEEE}
\thanks{The authors are with the Lab for Video and Image Analysis (LFOVIA), Department
of Electrical Engineering, Indian Institute of Technology Hyderabad, Yeddumailaram,
India, 502205 e-mail: \{ee12p1002, ee13m14p100001, sumohana\}@iith.ac.in.}}
\maketitle
\begin{abstract}
The tremendous growth in 3D (stereo) imaging and display technologies has led to 
 stereoscopic content (video and image) becoming increasingly popular. However, both the subjective and the objective evaluation of stereoscopic video content has not kept pace with the rapid growth of the content. Further, the availability of standard stereoscopic video databases is also quite limited. In this work, we attempt to alleviate these shortcomings. We present a stereoscopic video database and its subjective evaluation. 
We have created a database containing a set of 144 distorted videos. We limit our attention to H.264 compression artifacts. The distorted videos were generated using 6 uncompressed pristine videos of left and right views originally created by Goldmann et al. at EPFL \cite{goldmann2010comprehensive}. Further, 19 subjects participated in the subjective assessment task. Based on the subjective study, we have formulated a relation between the 2D and stereoscopic subjective scores as a function of compression rate and depth range. We have also evaluated the performance of popular 2D and 3D image/video quality assessment (I/VQA) algorithms on our database.
\end{abstract}

\IEEEpeerreviewmaketitle

\section{INTRODUCTION}
\label{sec:intro}

With the rapid advancements in 3D video technology, the industry and consumer experiences are improving in a tremendous way. According to the recent survey by Motion Picture Association of America (MPAA) \cite{mpaa2014}, the US revenue from 3D film industry has risen to 16\% in  2013 and one third of movie lovers  watch at least one 3D movie in a month. The primary reason for this incredible increase could be attributed to the depth-enhanced viewing experience. 
This has led to the movie and gaming industries investing a significant amount of resources on the creation of 3D content. 

The creation of multimedia content happens over several processing stages (such as sampling, quantization, demosiacing etc.), each of which could potentially degrade the perceptual quality of the content.
Compression artifacts are a very common cause of quality degradation. In this work, we focus our attention only on compression artifacts. 
Given that most of this content is meant for human consumption, the most relevant and consistent method to evaluate video quality is via subjective assessment. 
While subjective assessment is cumbersome, expensive and time consuming, this data is very essential to test the performance of objective VQA algorithms. 

In this paper, we present the subjective quality assessment of stereoscopic videos. The subjective assessment for stereoscopic videos is different from that of the 2D video, as the stereoscopic video consists of two views: left view and right view. These two views contribute to the perception of depth. Therefore, the overall quality of a stereoscopic video is a function of the individual qualities of the constituent left and right views.

Goldmann et al. \cite{goldmann2010comprehensive} created a database to study the effect of variation in the distance between the camera and the objects on perception. Ha et al. \cite{ha2011perceptual} performed a subjective study on a stereoscopic video data set based on the consideration of visual quality, depth perception, visual comfort and overall quality. Their work mainly focused on the perception of depth information without considering the distortion in the videos. Hewage et al. \cite{hewage2013study} conducted a subjective study to explore the effect of random packet loss artifacts on the overall perceptual quality of stereoscopic video. Aflaki et al. \cite{aflaki2010subjective} have performed a subjective study to explore the effects of asymmetric encoding (different rates and resolutions assigned to the left and right views) of a stereoscopic video. They conclude that asymmetric encoding offers bitrate savings compared to the symmetric case. 
They also concluded that PSNR is not a good objective measure for analyzing blocking artifacts and blurriness. Urvoy et al. \cite{urvoy2012nama3ds1} created a symmetrically distorted stereoscopic video dataset composed of H.264, JPEG 2000 as compression artifacts. However, this database does not consider the important case of asymmetric distortion of the stereoscopic views. While these subjective studies have considered either symmetric or asymmetric distortions, they have not considered the relationship between stereoscopic views and 2D views as a function of compression rate and depth.


While our work is similar in philosophy to \cite{aflaki2010subjective}, \cite{urvoy2012nama3ds1} we would like to highlight our contributions: a) creation of a stereoscopic video database that would be made freely available to the research community, b) a study of the relationship between the stereoscopic subjective scores and 2D subjective  scores, c) a performance evaluation of popular 2D and 3D I/VQA algorithms on our database, and d) an exploration of the effect of depth and compression rate on the perceptual quality of stereoscopic video.


The rest of the paper is organized as follows: Section \ref{sec:Database Explanation} explains the database and Section  \ref{sec:subjective study} describes the subjective assessment. Section \ref{sec:Subjective Scores Analysis} explains the analysis of subjective scores followed by a description of performance evaluation in Section \ref{sec:Performance Evaluation}. The conclusions are presented in Section \ref{sec:conclusion}.

\section{Database Description}
\label{sec:Database Explanation}
In this section we describe the generation of the video sequences used in our study, starting with a description of the pristine or reference sequences.
\subsection{Pristine Sequences}
\begin{figure}[H]
\centering
\begin{subfigure}[b]{0.15\textwidth}
\includegraphics[width=2.3cm,height=1.2cm]{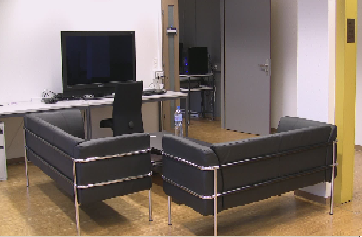}
\subcaption{\small Sofa.}
\end{subfigure}
\begin{subfigure}[b]{0.15\textwidth}
\includegraphics[width=2.3cm,height=1.2cm]{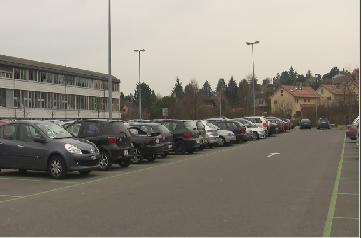}
\subcaption{\small Bike.} 
\end{subfigure}
\begin{subfigure}[b]{0.15\textwidth}
\includegraphics[width=2.3cm,height=1.2cm]{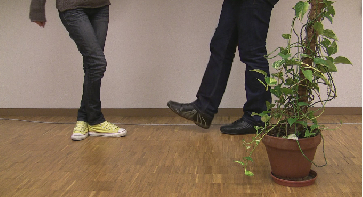}
\subcaption{\small Feet.} 
\end{subfigure}
\\
\begin{subfigure}[b]{0.15\textwidth}
\includegraphics[width=2.3cm,height=1.2cm]{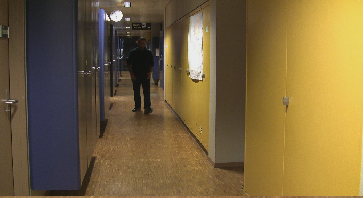}
\subcaption{\small Hallway.}
\end{subfigure}
\begin{subfigure}[b]{0.15\textwidth}
\includegraphics[width=2.3cm,height=1.2cm]{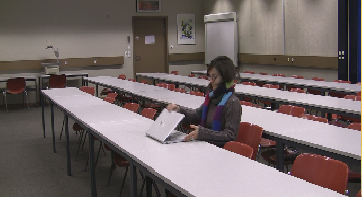}
\subcaption{\small Notebook.} 
\end{subfigure}
\begin{subfigure}[b]{0.15\textwidth}
\includegraphics[width=2.3cm,height=1.2cm]{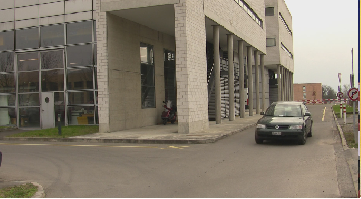}
\subcaption{\small Car.} 
\end{subfigure}
\caption{One frame from each pristine video.}
\label{videoframe}
\end{figure}

Goldmann et al. at EPFL \cite{goldmann2010comprehensive} created an open source database to highlight the effect of viewing distance variation between the camera and objects. We used the same reference videos as those in the EPFL study. 
There are six pristine videos per view (left view, right view) in the database. Fig. \ref{videoframe} shows one frame of each reference sequence in the database.
 This database consists of a collection of indoor and outdoor scenes with varying range of color, texture and objects. These videos are captured with identical camcorders placed horizontally with the separation continuously adjustable in the range 7--50 cm. The videos have resolutions varying from 1836$\times$1056 to 1900$\times$1054 pixels and a frame rate of 25 fps. Each video is 10 seconds in duration and is placed in an {\em{avi}} container. The camcorders were controlled by a remote to account for any temporal mismatch. 

 We grouped the 6 videos into two categories based on the depth content in them. {\em{Sofa, Feet, Hallway, Notebook}} sequences form group I having lower depth (3m - 10m). The {\em{Bike}} and {\em{Car}} sequences fall into group II having higher depth range ($>$ 100m).
\subsection{Test Sequences}
The pristine sequences that were in the {\em{avi}} format were converted to the YUV 4:2:0 format using the open-source {\em{ffmpeg}} application \cite{website:ffmpeg}. 
\begin{figure}[H]
\centering
\begin{subfigure}[b]{0.15\textwidth}
\includegraphics[width=3.2cm,height=1.5cm]{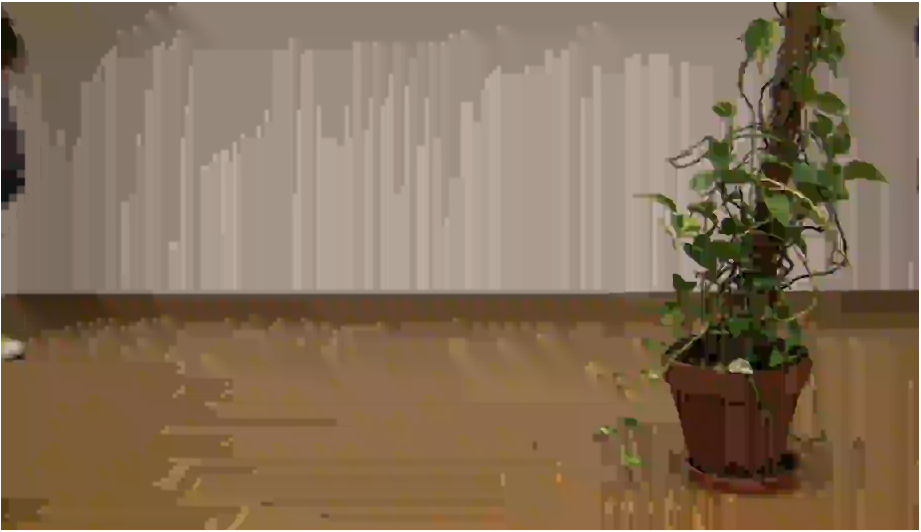}
\subcaption{\small 100 kbps.}
\end{subfigure}
~
\hspace{0.5cm}
\begin{subfigure}[b]{0.15\textwidth}
\includegraphics[width=3.2cm,height=1.5cm]{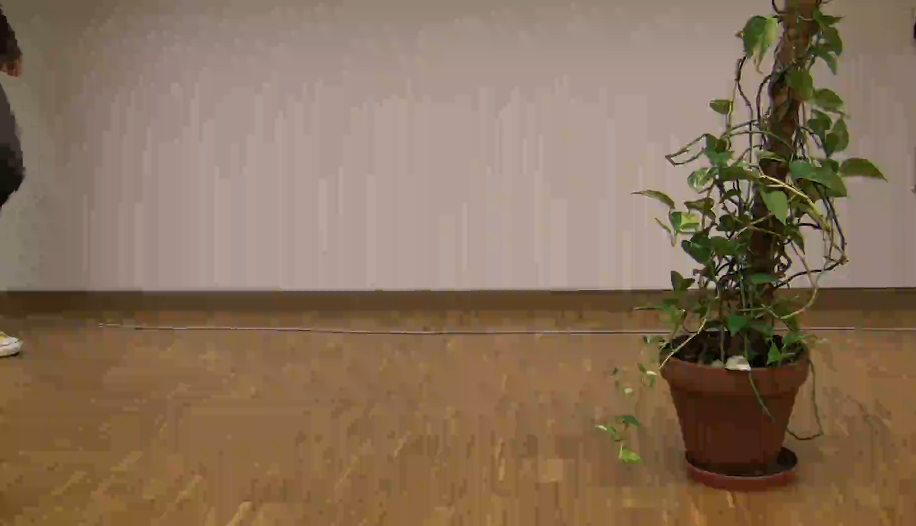}
\subcaption{\small 200 kbps.} 
\end{subfigure}
\\
\begin{subfigure}[b]{0.15\textwidth}
\includegraphics[width=3.2cm,height=1.5cm]{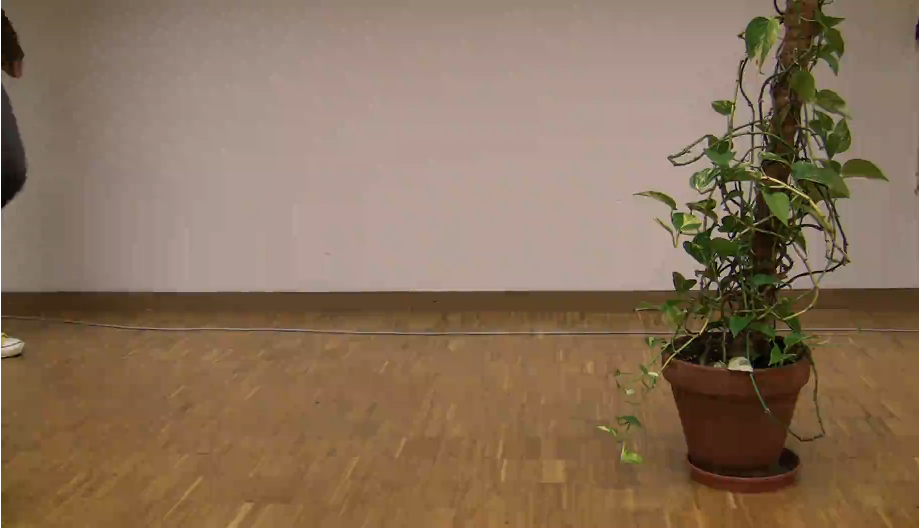}
\subcaption{\small 350 kbps.}
\end{subfigure}
~
\hspace{0.5cm}
\begin{subfigure}[b]{0.15\textwidth}
\includegraphics[width=3.2cm,height=1.5cm]{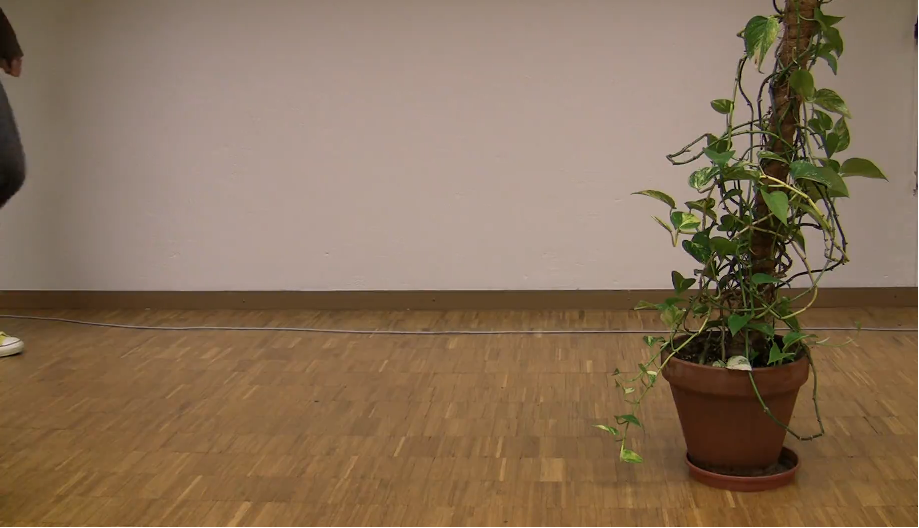}
\subcaption{\small 1200 kbps.} 
\end{subfigure}
\caption{One frame from the {\em{Feet}} sequence for different compression rates.}
\label{fig:distortedframe}
\end{figure}

As mentioned previously, the pristine videos had different resolutions, of which a majority were at a resolution of 1836$\times$1056 pixels. To maintain consistency, videos at other resolutions were resized to 1836$\times$1056 pixels using {\em{ffmpeg}}.
 
We generated 4 test sequences from each reference video using H.264 compression. We used a variety of compression rates (100 kbps, 200 kbps, 350 kbps and 1200 kbps) to cover a wide range of possible video transmission link rates. We capped our rate at 1200 kbps because the perceptual quality variation was not significant beyond this rate. Fig. \ref{fig:distortedframe} shows the frames of the pristine {\em{Feet}} sequence encoded at varying compression rates.  The compression was done by the {\em{ffmpeg}} software using {\em{libx264}} at the following settings: GOP length of 250 frames (default), CABAC encoder, flags and loop filter enabled. The compression rate was fixed using the {\emph{maxrate}} parameter. 

Overall, there are 24 test sequences and 6 reference sequences per view. The left and right views were combined to form symmetric and asymmetric sets. We would like to recall that in the symmetric set, the left and right views have been encoded at the same bitrate. In the asymmetric set, the left and right views of a video are encoded at
different bitrates. The symmetric set contains 24 videos (6 reference videos encoded at 4 different rates). The asymmetric set has 120 videos (out of the 150 possible permutations, 24 belong to the symmetric set, 6 are reference pairs, and the remaining fall into the asymmetric set). 
\section{SUBJECTIVE STUDY}
\label{sec:subjective study}
\subsection{Display Settings}
\label{sec:Display}
\begin{figure}[ht]
  \centering
    \includegraphics[width=0.28\textwidth]{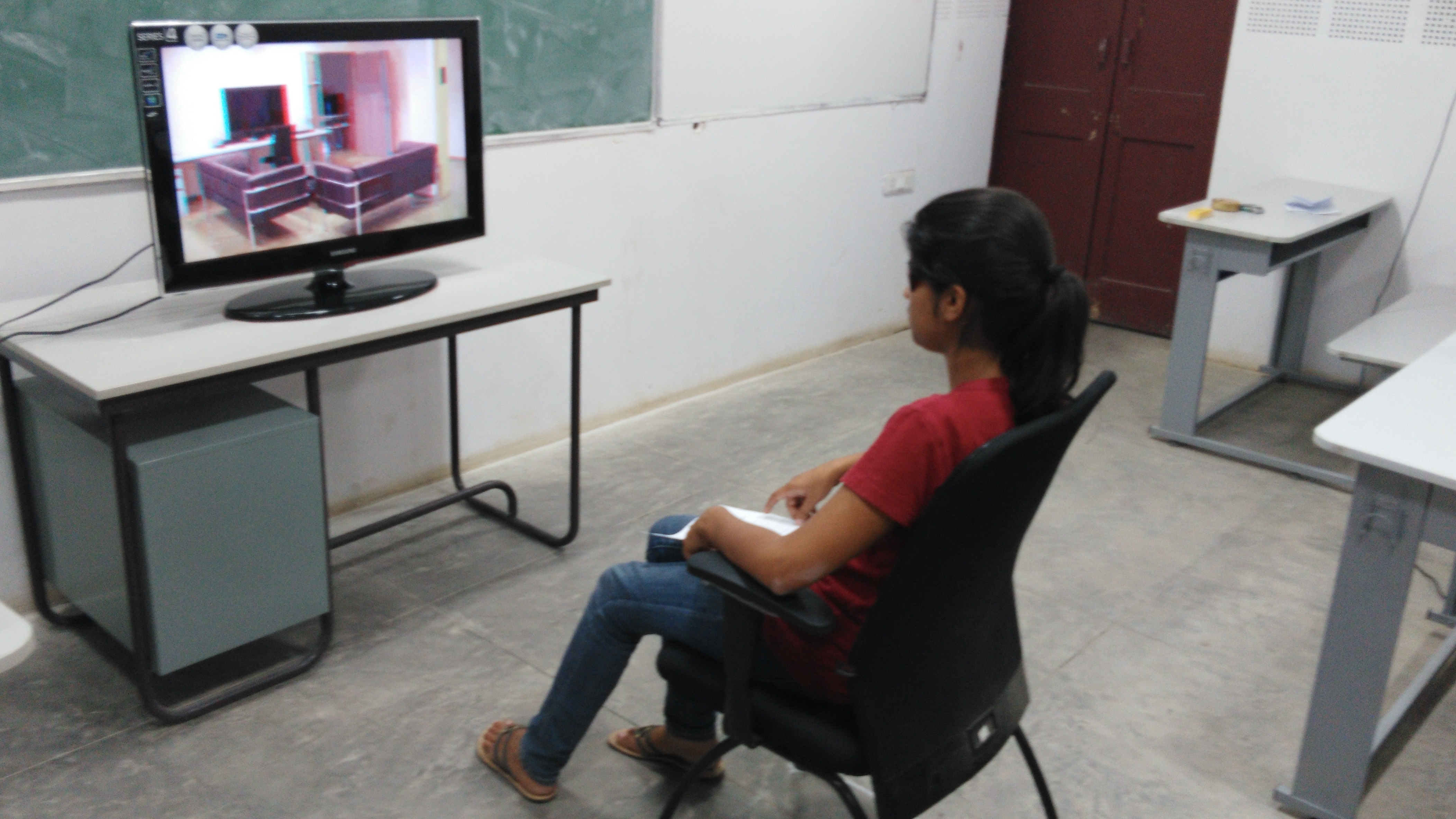}
    \caption{Subjective equipment setup.}
    \label{fig:setup}
  \end{figure}

We used a Samsung display of 32” inches (81.28 cm) with a screen resolution of 1366 $\times$ 768 pixels for our subjective study. The distance between the observer and screen was fixed at 1.5 meters which is 3 times the height of screen and the observer was seated at a height of 20.5” inches (52 cm) as shown in Fig. \ref{fig:setup}. The rest of the settings adhered to the ITU-R recommendations for subjective quality evaluation \cite{ITU}. The stereoscopic videos were played using a NVIDIA stereoscopic player \cite{website:NVIDIA}.
\subsection{Assessment Method}
\label{sec:Assessment Method}
We used the Single Stimulus Continuous Quality Evaluation (SSCQE) method to obtain the subjective rating of the videos. 
Our subjective study involved 19 subjects, gender distribution was not limited in our study and the average age of all observers is 24 years.
 A demo sequence that is representative of the quality variability in the distorted videos was first shown to the subjects.
  The subjective analysis was conducted in two sessions of 30 minutes each. During the first session the subjects were shown the left and right views 
of the 2D video. The videos were arranged in a random order of varying compression rates, and it was ensured that there were no repetition of video sequences. In the second session, the subjects were trained to perceive the stereoscopic content and asked to rate the stereoscopic videos. For stereoscopic quality evaluation, the subjects
wore a pair of anaglyph glasses and the stereo videos were rendered using a NVIDIA stereoscopic player.

The subjective rating given to the video is according to the ITU-R ACR scale, which ranges from 1 - 5 (1 - bad, 2 -poor, 3 - fair, 4 - good, 5 - Excellent). Non-integer ratings were also allowed. 
\section{Subjective Scores Analysis}
\label{sec:Subjective Scores Analysis}
\subsection{Subjective data handling}
To process the subjective scores we followed the ITU-R recommendations \cite{ITU}\cite{seshadrinathan2010study}. We have 150 (symmetric + asymmetric) scores for a stereoscopic video set and 30 scores for each 2D view (left view and right view). First, we compute difference scores between the test video and reference video. These scores are computed by subtracting the quality score assigned by the subject to a test video from the quality score assigned by the same subject to the corresponding reference video. 
\begin{equation}
d_{ij} = s_{ij_{ref}}-s_{ij}, \label{eq1}
\end{equation}
where 
${i}$ indicates the subject and ${j}$ indicates the video sequence id. The difference scores for the reference videos are not considered for analysis. 
The $z$-scores are computed by calculating the mean ($\mu_{i}$) and standard deviations ($\sigma_{i}$) from difference scores for each subject. 
The $z_{ij}$ scores are given by
\begin{equation}
\mu_{i}=\frac{\sum_{j=1}^{N_{j}} d_{ij}} {N_{j}}, \label{eq2}
\end{equation}
\begin{equation}	
\sigma_{i}=\sqrt{\frac{\sum_{j=1}^{N_{j}} (d_{ij}-\mu_{i})^2} {N_{j}-1} }, \label{eq3}
\end{equation}
\begin{equation}
z_{ij} = \frac{d_{ij}-\mu_{i}}{\sigma_{i}}, \label{eq4}
\end{equation}
where $N_j$ is the number of videos rated by the subject $i$. For the  stereoscopic case, $N_j = 150$ and for the 2D cases, $N_j = 30$ for each view. 

To remove outliers we followed the ITU-R BT 500.11 recommendations for observer screening. Observers are discarded if they exhibit a strong shift of votes compared to the average behaviour. 
In our analysis no outliers were found.
 
The $z$-scores lie in the range of [-3,3] which was scaled to [0,100] by
\begin{equation}
{zs}'_{ij} = \frac{100(z_{ij}+3)}{6},
\end{equation}
The final step in subjective processing is calculation of DMOS scores. DMOS is calculated by taking the mean of the rescaled $z$-scores across all the subjects per video. 
\begin{equation}
\text{DMOS}_{j}=\frac{\sum_{i=1}^{M} zs'_{ij}} {M},
 \label{eq6}
\end{equation}
where $M=19$. The range of DMOS values obtained for stereoscopic set is [79.73 28.9]. Similarly, for the left video set the range is [73.99 26.47] while it is [72.36 28.65] for the right video set.
\section{Performance Evaluation}
\label{sec:Performance Evaluation}
\subsection{Subjective Score based Evaluation}
Let $L_j$, $R_j$ be the DMOS for the left and right views respectively for a video $j$. The average of the left and right view DMOS, $V_j$ is given by
\begin{equation}
V_j = \frac{L_j+R_j}{2}.
\end{equation}
Table \ref{table:2D_stero} shows the correlation between $V_j$ of a video with the corresponding stereoscopic DMOS. 
As defined earlier, in the symmetric case both views having same compression rate while the asymmetric case stands for different compression rates in the left and right view. 
For instance, the asymmetric case for 100 kbps compression rate denotes the compression rate of one of the views being fixed at 100 kbps and the other view's rate being varied for all combinations and vice versa. 
\begin{table}[htbp]
\caption{Correlation between the the average DMOS of left and right views $V_j$ and the stereoscopic DMOS across varying compression rates} 
\centering 
\begin{tabular}{|c| c| c| } 
\hline
 \bf Compression rates &  \bf Symmetric  &  \bf Asymmetric  \\
\hline
\bf 100 kbps & 0.280 & 0.563 \\
\hline
\bf 200 kbps & 0.939 &  0.912 \\
\hline 
\bf 350 kbps & 0.872 &  0.934 \\
\hline
\bf 1200 kbps & 0.081 &  0.875 \\
\hline
\end{tabular}
\label{table:2D_stero} 
\end{table}
From Table \ref{table:2D_stero} it is clearly seen that $V_j$ is not a representative of the stereoscopic quality across the compression rates. 
Table \ref{table:Asymm} shows the correlation values between $V_j$ and the stereoscopic DMOS of group I (lower depth range) and group II (higher depth range) for different compression rates.
\begin{table}[htbp]
\caption{Correlation scores of average DMOS, $V_j$ and the stereoscopic DMOS as a function of depth and compression rates for videos belonging to group I and group II.} 
\centering 
\begin{tabular}{|c| c| c|} 
\hline
 \multirow{2}{*}{\bf Compression rates} &  \multicolumn{2}{c|} {\bf Asymmetric}  \\
 \cline{2-3}
 & \bf I &  \bf II  \\
 \hline
\bf 100 kbps & 0.48  & 0.62\\
\hline
\bf 200 kbps &0.77 & 0.92 \\
\hline 
\bf 350 kbps & 0.82 &  0.96 \\
\hline
\bf 1200 kbps & 0.77  & 0.94 \\
\hline
  \end{tabular}
\label{table:Asymm} 
\end{table}
\begin{table}[htbp]
\caption{Comparison of stereoscopic DMOS for asymmetric video set of group I videos (lower depth range).} 
\centering 
\begin{tabular}{|c| c| c| c| c| } 
\hline
 \multirow{2}{*}{\bf Compression rates} & \multicolumn{4}{c|}{\bf Asymmetric}\\
 \cline{2-5} 
 & \bf 100 kbps  &  \bf 200 kbps  &  \bf 350 kbps  &  \bf 1200 kbps \\
\hline
\bf 100 kbps &  & 65.06 & 66.16 & 59.06 \\
\hline
\bf 200 kbps & 70.87 &  & 54.45 & 48.29 \\
\hline 
\bf 350 kbps & 66.88 & 52.18 &  & 36.29 \\
\hline
\bf 1200 kbps & 62.05 & 49.25 & 42.65 & \\
\hline
\end{tabular}
\label{table:Asymm_I} 
\end{table}
\begin{table}[htbp]
\caption{Comparison of stereoscopic DMOS for asymmetric sets of group II videos (higher depth range).} 
\centering 
\begin{tabular}{|c| c| c| c| c| } 
\hline
  \multirow{2}{*}{\bf Compression rates} & \multicolumn{4}{c|}{\bf Asymmetric}\\
 \cline{2-5} 
 &  \bf 100 kbps  &  \bf 200 kbps  &  \bf 350 kbps  &  \bf 1200 kbps \\
\hline
\bf 100 kbps &  & 56.89 & 57.31 & 54.04 \\
\hline
\bf 200 kbps & 63.27 &  & 45.01 & 45.03 \\
\hline 
\bf 350 kbps & 61.71 & 44.22 &  & 35.78 \\
\hline
\bf 1200 kbps & 59.88 & 43.16 & 36.5 & \\
\hline
\end{tabular}
\label{table:Asymm_II} 
\end{table}
\begin{figure*}[htbp]
\centering
\begin{subfigure}[b]{0.22\textwidth}
\includegraphics[width=1\textwidth]{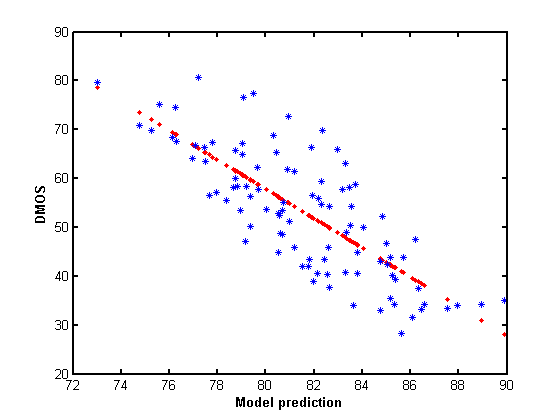}
\subcaption{PSNR \cite{sheikh2006statistical}.}
\end{subfigure}
\begin{subfigure}[b]{0.22\textwidth}
\includegraphics[width=1\textwidth]{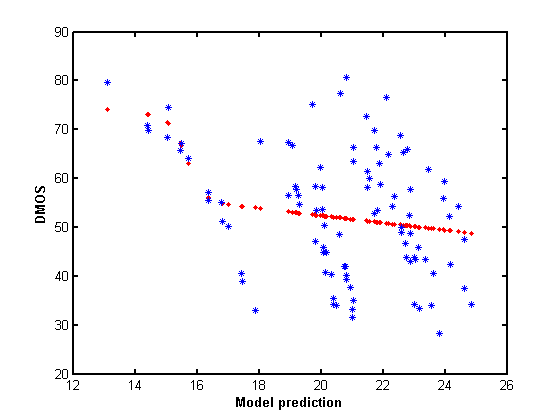}
\subcaption{VSNR \cite{chandler2007vsnr}.}
\end{subfigure}
\begin{subfigure}[b]{0.22\textwidth}
\includegraphics[width=1\textwidth]{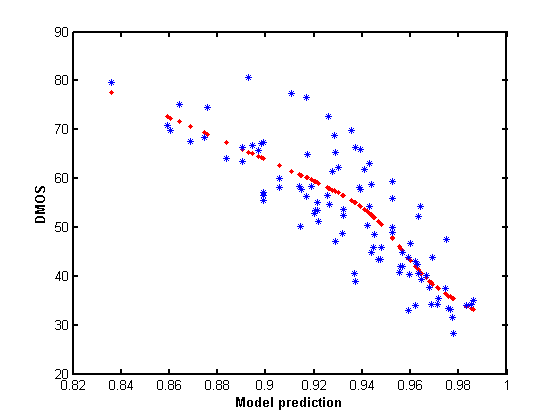}
\subcaption{SSIM \cite{Wang2004}.}
\end{subfigure}
\begin{subfigure}[b]{0.22\textwidth}
\includegraphics[width=1\linewidth]{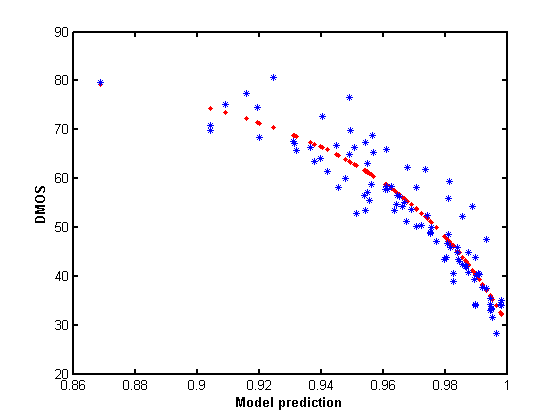}
\subcaption{FSIM \cite{zhang2011fsim}.}
\end{subfigure}
\\
\begin{subfigure}[b]{0.22\textwidth}
\includegraphics[width=1\linewidth]{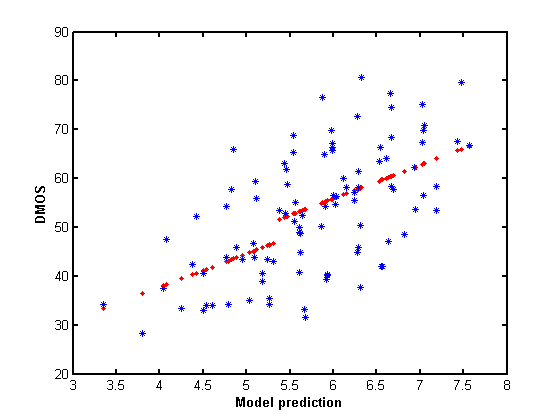}
\subcaption{STMAD \cite{vu2011spatiotemporal}.}
\end{subfigure}
\begin{subfigure}[b]{0.22\textwidth}
\includegraphics[width=1\linewidth]{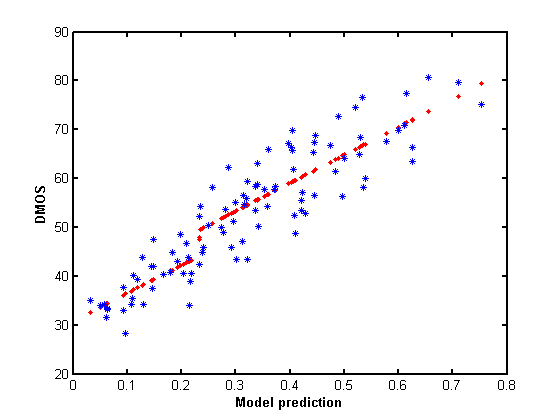}
\subcaption{BVQM \cite{website:vqm, pinson2004new}.}
\end{subfigure}
\begin{subfigure}[b]{0.22\textwidth}
\includegraphics[width=1\linewidth]{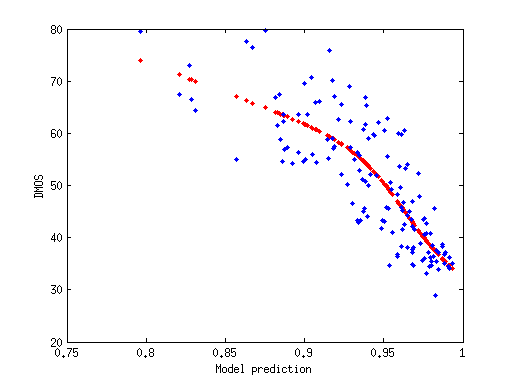}
\subcaption{Chen et al. \cite{chen2013full}.}
\end{subfigure}
\begin{subfigure}[b]{0.22\textwidth}
\includegraphics[width=1\linewidth]{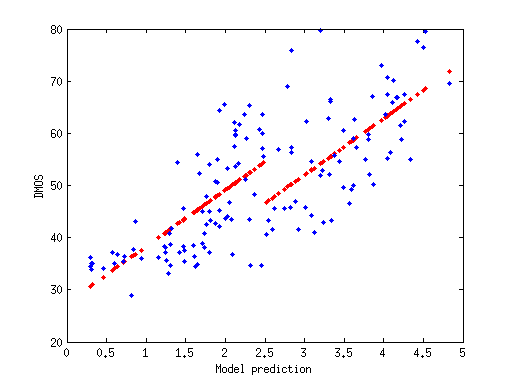}
\subcaption{STRIQE \cite{khan2015full}.}
\end{subfigure}
\caption{Scatter plots of objective scores versus the stereo DMOS over all the videos in the database.}
\label{scatterplots}
\end{figure*}
We present the following hypothesis to explain the performance of $V_j$ as a stereoscopic quality metric.
When a video is visually very annoying the viewer gets accustomed to it and tries to extract information from the given quality video. In case of stereoscopic video this information can be the depth range. As $V_j$ does not have the effect of depth in it (since it the average of 2D scores), the correlation is low for higher compression rates. However, in the case of medium compression rate, the scenes are neither visually too annoying nor very good and hence results in viewer dilemma. Owing to this constraint, the viewer does not attempt to infer additional information at this rate. 
Therefore, as the compression rate decreases the correlation increases. When the video is of very high quality (lower compression rate), the viewer tries to capture additional information from the scene which is again depth in the stereoscopic case, and hence the correlation decreases. 

\begin{table}[h]
\caption{Comparison of the performance of the 2D metrics on the left and right views of the database across the compression rates.} 
\centering 
\begin{tabular}{|c| c| c| c| c| } 
\hline
  \multirow{2}{*}{\bf Algorithm} & \multicolumn{4}{c|}{\bf Compression rates}\\
 \cline{2-5} 
  &  {\bf 100 kbps}  & {\bf 200 kbps} & {\bf 350 kbps} &{\bf 1200 kbps}  \\
\hline
PSNR \cite{sheikh2006statistical} & 0.64 & 0.53 & 0.29 & 0.76  \\
\hline
VSNR \cite{chandler2007vsnr} & 0.19 & 0.77 & 0.11 & 0.125  \\
\hline
SSIM \cite{Wang2004} & 0.69 & 0.73 & 0.45 & 0.77  \\
\hline
FSIM \cite{zhang2011fsim} & 0.69 & 0.78 & 0.49 & 0.76   \\
\hline
STMAD \cite{vu2011spatiotemporal} & 0.27 	& 0.42 & 0.62 & 0.72 \\
\hline
BVQM \cite{website:vqm, pinson2004new} & 0.74 & 0.93 & 0.89 & 0.76  \\
\hline
\end{tabular}
\label{table:2D_metric_corr}
\end{table}

\begin{table}[htbp]
\caption{Comparison of the performance of the average of the 2D objective metrics on the left and right views of the stereoscopic database across the compression rates for group I (lower depth range) and group II (higher depth range videos) - Linear Correlation Coefficient.} 
\centering 
\begin{tabular}{|c| c| c| c| c| c| c| c| c|} 
\hline
  \multirow{2}{*}{\bf Algorithm} &  \multicolumn{8}{c|}{\bf Asymmetric}\\
  \cline{2-9}
  &\multicolumn{8}{c|}{\bf Compression rates}\\
 \cline{2-9} 
  &  \multicolumn{2}{c|}{\bf 100 kbps}  & \multicolumn{2}{c|}{\bf 200 kbps} & \multicolumn{2}{c|}{\bf 350 kbps} &\multicolumn{2}{c|}{\bf 1200 kbps} \\
  \cline{2-9}
 & \bf I & \bf II & \bf I & \bf II & \bf I & \bf II & \bf I & \bf II  \\
\hline
PSNR \cite{sheikh2006statistical} & 0.71 & 0.55 & 0.63 & 0.74 & 0.70 & 0.37 & 0.68 & 0.64  \\
\hline
VSNR \cite{chandler2007vsnr} & 0.66 & 0.45 & 0.54 & 0.71 & 0.57 & 0.68 & 0.53 & 0.50   \\
\hline
SSIM \cite{Wang2004} & 0.71 & 0.86 & 0.77 & 0.90 & 0.85 & 0.96 & 0.81 & 0.95  \\
\hline
FSIM \cite{zhang2011fsim} & 0.97 & 0.94 & 0.95 & 0.96 & 0.92 & 0.92 & 0.98 & 0.98  \\
\hline
BVQM \cite{website:vqm, pinson2004new} & 0.73 & 0.74 & 0.88 & 0.92 & 0.92 & 0.96 & 0.94 & 0.97  \\
\hline
STMAD \cite{vu2011spatiotemporal} & 0.66 & 0.69 & 0.75 & 0.89 & 0.77 & 0.90 & 0.80 & 0.82  \\
\hline
\em{Chen et al.} \cite{chen2013full} & 0.62 & 0.92 & 0.64 & 0.98 & 0.77 & 0.98 & 0.88 & 0.97 \\
\hline 
\em{STRIQE} \cite{khan2015full}& 0.65 & 0.77 & 0.71 & 0.89 & 0.77 & 0.91 & 0.74 & 0.87 \\
\hline
 \end{tabular}
\label{table:2D_metric_stereo} 
\end{table}



Tables \ref{table:Asymm_I} and \ref{table:Asymm_II} show the average stereoscopic DMOS values for the asymmetric video sets for groups I and II respectively. It is clear that the stereoscopic DMOS for the group I videos are high compared to the group II videos. In group II videos the depth range is high which results in lower viewing precision of the objects in the scene. Therefore, the distortions at higher depth range are not easily perceived. Thus, for a given compression rate, we can conclude that the DMOS for stereoscopic videos with higher depth range is always less than the videos with lower depth range.
\subsection{Objective Score based Evaluation}
In order to test the efficacy of popular 2D and 3D objective I/VQA metrics on stereoscopic video, they were evaluated on the stereoscopic database we have created. Standard measures of performance such as Spearman Rank Order Correlation Coefficient (SROCC) and Linear Correlation Coefficient (LCC) were used.
Table \ref{table:2D_metric_corr} shows the performance of the 2D I/VQA metrics on the left and right view videos of the database.
 A non-linear regression on the VQA scores is done using the logistic function mentioned in \cite{website:LIVE_Database_Report} and LCC is computed between the fitted objective scores and the DMOS. PSNR \cite{sheikh2006statistical}, VSNR \cite{chandler2007vsnr}, SSIM \cite{Wang2004}, FSIM \cite{zhang2011fsim} are image metrics and they are applied on a frame by frame basis and averaged. ST-MAD \cite{vu2011spatiotemporal} and BVQM \cite{website:vqm, pinson2004new} are 2D video metrics. The Chen et al. \cite{chen2013full} and STRIQE \cite{khan2015full} are 3D IQA metrics. Table \ref{table:2D_metric_stereo} illustrates the performance of the 2D and 3D I/VQA metrics across compression rates for stereoscopic videos in the database.  
\section{CONCLUSIONS AND FUTURE WORK}
\label{sec:conclusion}
The purpose of this study was to create a H.264 compressed stereoscopic video dataset. The created stereoscopic video database composed of 144 videos was created using the 6 pristine videos from the EPFL database \cite{goldmann2010comprehensive}, and compressed at 4 compression rates. The subjective study was done on these videos by 19 subjects. 
We tested the efficacy of several 2D I/VQA algorithms on the proposed database.

From the analysis of the subjective scores we made the following conclusions: i) the average DMOS from the left and right views $V_j$ is not a representative of the stereoscopic DMOS, ii) depth plays a role at very high and very low compression rates. Therefore, the 2D and stereoscopic I/VQA algorithms do not perform well at high and low compression rates, iii) the study on the correlation values has depicted that at a given compression rate, the videos with higher depth range have better visual quality compared to that of lower depth range ones. The objective VQA perform better on the videos having higher depth range. 

We plan to make the database and the DMOS values available publicly to the research community. 

\appendices
\ifCLASSOPTIONcaptionsoff
  \newpage
\fi

\bibliographystyle{ieeetr}
\bibliography{master_ref}
\end{document}